\documentclass[12pt]{article}

\usepackage{epsfig}
\usepackage{graphicx}

\title{The quantum geometric limit}

\author
{Seth Lloyd\\
\\
\normalsize{Department of Mechanical Engineering}\\
\normalsize{MIT 3-160, Cambridge MA 02139 USA}\\
\\
\normalsize{slloyd@mit.edu}
}

\date{}

\begin{document}


\baselineskip23pt


\maketitle

\begin{abstract}
In Einstein's {\it gedankenexperiment} for measuring
space and time, an ensemble of clocks moving through curved 
spacetime measures geometry by sending signals back and forth, 
as in the global positioning system (GPS).  Combining well-known 
quantum limits to measurement with the requirement that the energy density of 
clocks and signals be be no greater than the black hole density leads
to the quantum geometric limit: the total number of ticks of
clocks and clicks of detectors that can be contained in a four volume
of spacetime of radius $r$ and temporal extent $t$ is
less than or equal to $rt/\pi \ell_P t_P$, where $\ell_P$, $t_P$
are the Planck length and time.   The quantum geometric limit
suggests that each event or `op' that takes place in a 
four-volume of spacetime is associated with a 
Planck-scale area.  This paper shows that the quantum geometric
limit can be used to derive general relativity:
if each quantum event is associated with a Planck-scale area 
{\it removed} from two-dimensional surfaces in the volume in which the event
takes place, then Einstein's equations must hold.  
\end{abstract}

The quantum geometric limit imposes a fundamental physical bound
to the accuracy with which quantum systems can measure the geometry of 
spacetime [1-2].  This limit arises naturally from the combination
of well-known quantum limits to measurement accuracy, together with 
the requirement that the apparatus used to measure space and time
be no denser than a black hole.  
The quantum geometric limit connects previous
limits to measuring spacetime [3-4], the physics of computation [5-7],
holography [8-13], and quantum mechanics on curved spacetime [14-15, 17].  
In particular, the holographic principle 
encourages us to imagine each bit within a spacelike three volume 
as projected onto the two-dimensional surface of that volume at 
a density of no greater than the Planck scale.  By contrast,
the quantum geometric limit encourages us to imagine each event or `op'
that occurs within a spacetime four volume as projected onto
the two-dimensional surfaces in that volume
at a density of no greater than the Planck scale.
In [15] Jacobson showed that the holographic area-entropy law can
be used as a basis for deriving Einstein's equations.
This paper shows that the same is true for the quantum
geometric limit: Einstein's equations follow from the assumption
that each quantum event or op {\it removes} a 
Planck-scale area from two-dimensional surfaces in the
volume in which the event occurs. 

\section{Limits to measuring spacetime}

The Margolus-Levitin theorem [5] states that the time $\Delta t$ it takes a
quantum system such as a clock to go from one state to an orthogonal state
is greater than or equal to $ \pi\hbar/2E$, where $E$ is
the expectation value of the energy of the system above the
ground state energy.  As an example of a system that saturates
the Margolus-Levitin theorem, consider a two-level system
or qubit with energy eigenstates $|0\rangle$ with energy $0$
and $|1\rangle$ with energy $\hbar\omega$.   The state
$|\psi(t)\rangle = (1/\sqrt 2)(|0\rangle + e^{-i\omega \Delta t} |1\rangle)$ 
then saturates the Margolus-Levitin theorem for $E=\hbar\omega/2$,
and $\Delta t = \pi\hbar/2E = \pi/\omega$.
The Margolus-Levitin theorem and its
variants such as the quantum speed limit [7] can be combined with
the fundamental physical limits to computation and measurement
[1-2, 5-7] to put bounds on the accuracy to which spacetime geometry
can be measured.  

Consider Einstein's seminal thought experiment
for measuring spacetime geometry, in which
spacetime is filled with a `swarm' of clocks that
map out the spacetime geometry by
exchanging signals with the other clocks and measuring the signals'
times of arrival.  
This thought experiment is of course the basis for the
global positioning system (GPS).  The clocks could be as large as
GPS satellites, or as small as elementary particles.  References
[1-2] put bounds on how accurately this swarm of clocks can map out
a sub-volume of spacetime with radius $r$ over time $t$. 
Every tick of a clock or click
of a detector is an elementary event in which a system goes from
a state to orthogonal state.  Accordingly, the the total number of ticks 
and clicks that can take place within the volume is limited by
the Margolus-Levitin theorem: it is less than or equal to
$ \# \equiv 2Et/\pi\hbar$, where $E$ is the quantum expectation value
of the energy of the clocks within
the volume, measured from their ground state.   
Call $\#$ the number of quantum `ops' accumulated
by the system.  A quantum op occurs when the state of the system
accrues an average phase $ Et=\pi/2$ 
relative to the phase of the ground state, whether the
system moves to an orthogonal state or not.  
For example, a thermal state of the system can be stationary but
still accrue quantum ops.  

If the clocks are packed too densely within some sub-volume, they
will form a black hole and be unable to send signals to clocks
outside their horizon.  That is, clocks within a black hole
cease to participate in global measurement of spacetime. 
(They may still measure spacetime within the horizon.)
To prevent the formation of an horizon, the total 
of clocks and signals within spacelike regions of radius $r$ 
must be less than $c^4r/2G$.  The Margolus-Levitin theorem
together with the requirement that the clocks and signals
participating in measuring spacetime be no greater than the black-hole density
implies the quantum geometric limit [1,2]:
the total number of elementary events and the number of ops
that can occur in such a volume of spacetime are bounded by
$$ \# \leq {c^4 rt \over \pi \hbar G} =  {rt\over \pi l_P t_P}. \eqno(1)$$ 
The quantum geometric limit (1) was derived without any recourse to
quantum gravity: the Planck scale makes its appearance simply
from combining quantum limits to measurement with the requirement
that sub-volumes of the GPS system be no denser than black holes.  

Rotating and charged black holes possess a more
complex Kerr-Newman structure [14,16,18] that could modify equation (1). 
Within a black hole, arbitrarily large
matter densities can occur in the approach to the
singularity, so clocks and signals could potentially 
violate equation (1).  
In a Schwarzschild black hole, the maximum proper time
ticked out by a free-falling clock after it passes the
event horizon at radius $R = 2GE/c^4$ is $ t = \pi R/c$ [18], suggesting
that the quantum geometric limit still holds in the sense
that the maximum number of ticks and clicks experienced by observers within the
hole before they hit the singularity is $\leq R^2/\ell_P^2$.  
The ability of clocks and signals falling into 
a complex singularity structure such as in Kerr-Newman
black holes which possess closed timelike curves [18] is an open question.   
In any case, the clocks and signals inside the 
black hole can not participate in the overall GPS system's measurement 
of spacetime outside the hole.  

Although straightforward to derive,
the quantum geometric limit's association of events with
two-dimensional Planck scale areas is somewhat surprising.
{\it A priori}, one
might have thought that the number of events within a
four volume would be limited by the measure of that volume   
divided by the Planck scale to the fourth power.
Alternatively, holography could be taken to suggest that the number of
events be proportional to the surface area of the volume
times time divided by Planck scale cubed.  The quantum
geometric limit (1) shows that the concentration of possible
events in spacetime is sparser than either of these 
guesses indicate.

\section{Deriving Einstein's equations from
the quantum geometric limit}

The holographic principle arose out of black-hole thermodynamics 
and quantum field theory on curved spacetime.
In [15], Jacobson turned the argument around: he showed 
how Einstein's equations can be
derived from combining the holographic entropy-area law
with the fact that accelerated observers see horizon
radiation with a temperature proportional to their acceleration.
(See also the work of Verlinde [19] and Dreyer [20].)

The quantum geometric limit suggests
each quantum event in spacetime is associated with a two-dimensional
Planck-scale area.  Now turn the argument around and derive Einstein's
equations from the quantum geometric limit.  The basic idea is straightforward: 
our {\it ansatz} is that
each elementary quantum `op' {\it removes} a Planck scale area 
from the two-dimensional sections of the spatial 
three volume in which the event occurs. 
The removal of area from a flat two-dimensional
section causes it to curve.  This curvature is the curvature
of spacetime, and as will now be shown, it induces the
spacetime to obey Einstein's equations.

Consider an inertial observer following a geodesic through spacetime.  Consider
a local region of spacetime in the vicinity of the observer sufficiently
small that the energy-momentum and curvature tensors are effectively
constant over this region, and the spacetime within the region
is close to Minkowski space (i.e., $K r^2 << 1$, where $r$ is the radius
of the region and $K$ is the maximum curvature within the region).  
The observer
describes this local region of spacetime by an orthonormal tetrad
of vectors (a vierbein) $\{{e^a}_{\mu}\}$.  $e^0_\mu$ is the
timelike tangent vector to her path, and ${e^j}_\mu$, $j=1,2,3$ are an
orthonormal triad of spacelike vectors.  We have
${e^a}_\mu {e^b}_{\nu} g^{\mu \nu}
= \eta^{ab}$, where $\eta^{ab} = {\rm diag}(-1,+1,+1,+1)$ is
the Minkowski-space metric, and $g^{\mu\nu}$ is the spacetime
metric.  Similarly, ${e^a}_\mu {e^b}_{\nu} \eta_{ab} = g_{\mu\nu}$.
Latin indices of ${e^a}_\mu$ are raised and lowered using $\eta_{ab}$,
and Greek indices are raised and lowered using $g_{\mu\nu}$.

The observer maps out her local region of spacetime using
GPS coordinates (figure 1).   Consider the causal diamond
$\diamondsuit(r,t)$ formed by the 
the intersection of the interiors of the forward
light cone emitted at time $t-r/c$ and the backward
light cone absorbed at time $t+r/c$.  
In Minkowski space the measure of the four-volume 
$\diamondsuit(r,t)$ is $2\pi r^4/3c$.
The light cones intersect at a 2-sphere $S$.  
Define the spacelike geodesic 2-disc $D_{12}$ by extending
from the observer's position at time $t$ the space-like geodesics 
initially tangent to the $12$ plane.
$D_{12}$ has circumference $C_{12}$ and Gaussian curvature $K_{12}$. 
Similarly, $D_{23}$ and 
$D_{31}$ are the geodesic 2-disks given by extending the
spacelike geodesics initially tangent to the
$23$ and $31$ planes.  These 2-disks have circumferences
$C_{23}$, $C_{31}$, and curvatures $K_{23}$, $K_{31}$.  

A quantum `op' corresponds to an average accumulation of
phase $E\Delta t/\hbar = \pi/2$ in the observer's frame, 
where $E$ is the observed energy above the local ground state energy.   
When a clock ticks
or a detector clicks, at least one quantum op is performed.   
The average energy density above the ground state energy
within the light cones as measured by the observer is $T^{\mu\nu}
{e^0}_{\mu} {e^0}_{\nu}$.  
The total number of ops recorded by the observer within 
the two light cones is therefore equal to 
$$ \# = (2/\pi \hbar) T^{\mu\nu} {e^0}_{\mu} {e^0}_{\nu} 
(2\pi/3) r^4/c. \eqno(3)$$

Our {\it ansatz} is that each op within the volume $\diamondsuit(r,t)$ 
removes a Planck-scale area
from the two-dimensional surfaces associated with the volume (figure 1). 
Compared with flat space, each op removes a net area 
$\Delta A = \alpha \ell_P^2 = \Delta A_{12} + \Delta A_{23} + 
\Delta A_{31}$ from the three 2-disks $D_{12}$, $D_{23}$,
$D_{31}$.  Removing area from an initially flat 2-disk causes 
it to curve. 
In the limit of small $r$ the Bertrand-Diquet-Puiseux theorem 
implies that the Gaussian curvature $K_{ij}$ of the $ij$ disk
of the disk is related to the deficit area $\Delta A_{ij}$ removed by 
$\Delta A_{ij} = \pi r^4 K_{ij}/12$.  Equivalently, the circumference
of the $ij$ disk is reduced by 
$ \Delta C_{ij} = \pi r^3 K_{ij}/3$, and the area 
of the two-sphere $S$ is reduced by $4\pi r^4 K/9$, where 
$K = K_{12} + K_{23} + K_{31}$ is the half the curvature of the spacelike
three volume orthogonal to the motion of the observer [21-22]. 
(I am indebted to T. Jacobson for pointing out the connection
to the deficit area of the 2-sphere [23].)

To relate the area removed to the Riemann tensor, note that
the Gaussian curvature of the geodesic 2-disk $D_{ij}$
is equal to the sectional curvature [21,24]:
$$K_{ij}= {R^\mu}_{\nu\rho\sigma} e_{i\mu} {e_j}^\nu {e_i}^\rho {e_j}^\sigma 
\eqno(4)$$
where ${R^\mu}_{\nu\rho\sigma}$ is the Riemann tensor.  
Removing a total area $\alpha \ell_P^2 \#$ from the three 2-disks
$D_{ij}$ then yields 
\begin{eqnarray}
\alpha \ell_P^2 \#  
&=& (\pi/12) ( K_{12} + K_{23} + K_{31}) r^4   \nonumber \\ 
&=& (\pi/24)(R^{\mu\nu} {e^0}_\mu {e^0}_\nu
+ \sum_{j=1}^3 R^{\mu\nu} {e^j}_\mu {e^j}_\nu) r^4  \nonumber \\  
&=& (\pi/12) (R^{\mu\nu} - (1/2) g^{\mu\nu} R) {e^0}_\mu {e^0}_\nu r^4.
\nonumber\quad\quad\quad(5)
\end{eqnarray}
Combining equation (5) for the change in area together with equation
(3) for the total number of ops yields
$$ (16/\pi \hbar) \alpha \ell_P t_P  T^{\mu\nu} {e^0}_{\mu} {e^0}_{\nu}
=
(R^{\mu\nu} - (1/2) g^{\mu\nu} R) {e^0}_\mu {e^0}_\nu.\eqno(6)$$
Setting $\alpha = \pi^2/2$, and noting that 
${e^0}_\mu$ can be any timelike unit vector -- that is,
equation (6) should hold for all observers --
implies Einstein's equations.  

Equations (5-6) show that if each op that takes place removes an
area $ A = \pi^2\ell_P^2$ from the initially flat two-dimensional
spacelike 2-disks $D_{ij}$ contained within the  
causal diamond $\diamondsuit(r,t)$ of intersecting light cones in which the op
takes place, inducing curvature, then the resulting curved
spacetime obeys Einstein's equations.  
Equivalently, Einstein's equations arise if each op
removes an area  $8\pi^2 \ell_P^2/3$ from the 2-sphere defined by the
intersection of the light cones.  Our derivation effectively 
`de-dimensionalizes' the observation [21-22] that in Einstein's equations
the local energy density as perceived by an observer is proportional
to the curvature of the spacelike 3-volume orthogonal to her path.
The derivation reveals the origin of the quantum geometric limit:
there is only so much area one can remove from a 2-sphere.

Note that this derivation yields
Einstein's equations without an intrinsic
cosmological constant.  The lack of an intrinsic cosmological
term arises from the requirement -- enjoined by the quantum
mechanics of measurement -- that the number of events or ops be an observable
quantity.  Consequently, that number can depend only on the energy above 
the ground state energy.  If only observable quantum phases contribute
to the interaction between matter and geometry, then
the vacuum energy does not contribute
to the gravitational energy.  Of course, the equations do not rule out a form
of matter that corresponds to a cosmological term.  

In certain circumstances -- e.g., at the horizon of a black hole, at
the mouth of a wormhole, or between the plates of capacitor (the
Casimir effect) -- quantum field theory allows the local energy density 
to be observably lower than the energy density of the global vacuum [14]. 
Does the possibility of such negative energy densities mean
that there can be negative ops (`nops')?  The definition
of an op should be able reconcile,
for example, the number of ops measured by an observer falling
into a black hole with that measured by an observer far from the hole.
A full quantum field theoretic treatment of the quantum geometric
limit lies outside of the scope
of the current paper, however, and will be undertaken elsewhere.

In general, the quantum state of the local matter 
is a mixture, e.g., a thermal state.  
Even when the overall state of the matter is pure, 
the local state is typically mixed due to entanglement: 
for example, vacuum entanglement is responsible for the entropy-area
law [13].   The Page-Geilker experiment [25] suggests that -- at least 
in relatively macroscopic situations --
different components of the mixture yield different local geometries.  
For a mixed state, the number
of ops can be evaluated separately in each component of the mixture. 
In an entangled state, the geometry induced by
a particular component of the local mixed state is correlated
with the geometry induced by the corresponding 
state of the matter elsewhere.  

Before closing, it is worthwhile to compare the derivation of Einstein's
equations from the quantum geometric limit with that of Jacobson [15].
Einstein's equations relate energy density to curvature.
Jacobson relates energy density to a dimensionless
quantity, entropy -- measured in bits --  
by introducing Planck's constant $\hbar$ in the
expression for the Unruh temperature. 
The entropy-area correspondence [13] suggests that
each bit of entropy is associated with a two-dimensional area $\eta$,
which Jacobson takes to be an area added to a
cross-section of the horizon.  The Raychaudhuri equation then implies 
Einstein's equations with gravitational constant $G=
\eta/(4\ln 2 \hbar)$, so that $\eta$ is indeed a Planck-scale area.

By comparison, the quantum geometric limit relates energy density 
to a dimensionless quantity -- number of ops -- by applying the
Margolus-Levitin theorem, which introduces $\hbar$.
The requirement that collections of clocks and signals within a four volume
not exceed the black-hole density
leads to the quantum geometric limit.  This limit 
suggests that each op is associated a two-dimensional Planck scale area,
which we take to be an area removed from two-dimensional surfaces within
the four volume in which the op takes place.  The introduction of the
Planck length squared inserts the gravitational constant and removes
$\hbar$.   The relationship between Gaussian curvature and sectional
curvature then implies Einstein's equations.

The main difference between the two approaches, of course, is that
in Jacobson's case the fundamental dimensionless quantity is a bit,
a unit of information, whereas here it is an op -- a unit of action
or change.

\section{Conclusion}

Intriguing connections between quantum information and gravity have
been arising for decades [1-4, 6, 8-13, 15-27].  This paper
attempted to elucidate those connections by applying
fundamental quantum limits to measurement of space and time. 
As in [15], the result is not a theory of quantum 
gravity {\it per se}, but rather a quantum theory which gives rise 
to general relativity under simple assumptions.
The quantum geometric limit states that the number of elementary
events such as clock ticks, detector clicks, or bit flips that
can be contained in a four volume of space time of covariant
radius $r$ and spatial extent $t$ is limited by $rt/\pi l_P t_P$.  
By bounding the number of ops that can take place within a
four volume, this limit is complementary to the holographic
limit, which bounds the number of bits associated with a spatial three
volume by the surface area of the volume divided by the Planck length
squared.  Holography encourages us to imagine the bits of
information characterizing the quantum state of the spatial 
three volume as projected onto the two-dimensional boundary of
the volume at a density no greater than on the order of one bit 
per Planck length
squared.   The quantum geometric limit encourages us to imagine the
elementary events or `ops' that occur within a spacetime four volume 
as projected onto two-dimensional surfaces in that
volume at a density no greater than on the order of one op per 
Planck length squared.  If each op removes a Planck-scale area from
those surfaces, then Einstein's equations hold.

\vfill
\noindent{\it Acknowledgements:} The author would like to thank O. Dreyer,
V. Giovannetti, T. Jacobson, and L. Maccone
for helpful discussions.  This work was supported by NSF, DARPA, Intel,
Lockheed Martin, ENI under the MIT Energy Initiative, ARO under a 
MURI program, and Jeffrey Epstein.

\vfil\eject

\noindent{\it References:}
\bigskip

\bigskip\noindent (1) V. Giovannetti, S. Lloyd, L. Maccone,
{\it Science} {\bf 306}, 1330 (2004); quant-ph/0412078.

\bigskip\noindent (2) S. Lloyd, Y.J. Ng, {\it Sci. Am.} {\bf 291}(5), 
52-61 (2004).

\bigskip\noindent (3) E.P. Wigner, Rev. Mod. Phys. 29, 255-268 (1957).

\bigskip\noindent (4) B. S. DeWitt, in {\it Gravitation: an 
Introduction to Current Research,} edited by L.  Witten 
(Wiley, New York, 1962). 

\bigskip
\noindent (5) Margolus, N., Levitin, L.B., in {\it PhysComp96}, T. Toffoli,
M. Biafore, J. Leao, eds. (NECSI, Boston) 1996; {\it Physica D} {\bf 120},
188-195 (1998).

\bigskip\noindent (6) S. Lloyd, {\it Nature} {\bf 406}, 1047-1054, 2000;
{\it Phys. Rev. Lett.} {\bf 88} (23): art. no. 237901, 2002. 

\bigskip\noindent (7) V. Giovannetti, S. Lloyd, L. Maccone, 
{\it Phys. Rev. A} {\bf 67}, 052109 (2003); {\it Europhys. Lett.} {\bf 62},
615 (2003); J. Opt. B: Quantum Semiclass. Opt. {\bf 6}, S807-S810 (2004).

\bigskip\noindent (8) J.D. Bekenstein, {\it Phys. Rev. D} {\bf 7},
2333 (1973). 

\bigskip\noindent (9) G. 't Hooft, gr-qc/9310026.

\bigskip\noindent (10) L. Susskind, {\it J. Math. Phys.} {\bf 36},
6377 (1995); hep-th/9409089.

\bigskip\noindent (11) R. Bousso, JHEP {\bf 7}, 4 (1999); hep-th/9905177.  

\bigskip\noindent (12) U. Yurtsever, {\it Phys. Rev. Lett.}{\bf 91}, 
041302, (2003). 

\bigskip\noindent (13) L. Bombelli, R.K. Koul, J. Lee, and R.D.
Sorkin, {\it Phys. Rev. D} {\bf 34}, 373-383 (1986).

\bigskip\noindent (14) N.D. Birrell, P.C.W. Davies, {\it Quantum
fields in curved space,} Cambridge University Press, Cambridge (1982).

\bigskip\noindent (15) T. Jacobson, {\it Phys. Rev. Lett.} {\bf 75},
1260-1263, (1995);
 arXiv:gr-qc/9504004.

\bigskip\noindent (16)
S.W. Hawking, G.F. Ellis, {\it The large-scale structure of space-time},
Cambridge University Press, Cambridge (1973).

\bigskip\noindent (17) A. Kempf, {\it Phys. Rev. Lett.} {\bf 103},
231301 (2009); arXiv:0908.3061.

\bigskip\noindent (18) G.F. Lewis, J. Kwan, {\it Pub. Astr. Soc. Aust.
} {\bf 24}, 46-52 (2007); arXiv: 0705.1029.

\bigskip\noindent (19) E. Verlinde,  {\it JHEP} {\bf 1104}, 029
(2011);  arXiv:1001.0785.

\bigskip\noindent (20) O. Dreyer,  arXiv:0710.4350,  arXiv:1203.2641.

\bigskip\noindent (21) W. Pauli, {\it Theory of relativity}, Dover, New
York (1958).

\bigskip\noindent (22) R.P. Feynman, F.B. Morinigo, W.G. Wagner,
{\it Feynman lectures on 
gravitation}, B. Hatfield, ed., Section 11.2, Addison-Wesley, Reading (1995).

\bigskip\noindent (23) T. Jacobson, to be published.

\bigskip\noindent (24) G.S. Hall, A.D. Rendall,
{\it Gen. Rel. Grav.} {\bf 19}, 771-789  (1987).

\bigskip\noindent (25) D.N. Page, C.D. Geilker, {\it Phys. Rev. Lett.}
{\bf 47}, 979-872 (1981).

\bigskip\noindent (26) J.D. Barrow,  {\it Phys. Rev. D}  {\bf 54}, 
6563-6564 (1996).

\bigskip\noindent (27) S. Lloyd, quant-ph/0501135.

\vfill\eject
\begin{figure}
\begin{center}
\includegraphics[height=500pt]{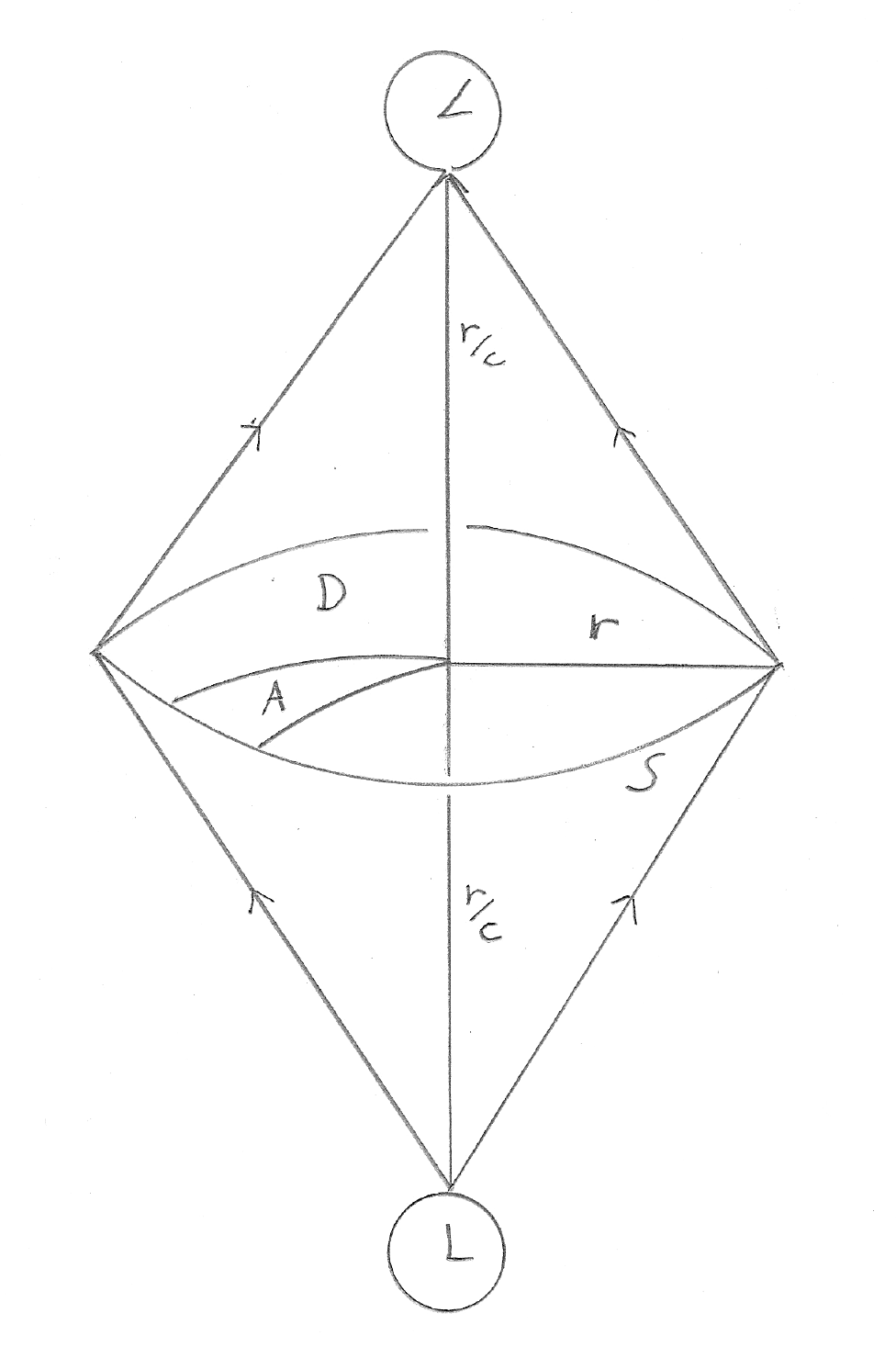}
\caption{
An inertial clock falling through spacetime sets up a local
coordinate system using GPS coordinates.  The 
forward light cone of signals emitted by the clock at time
$t-r/c$ intersects with the backward light cone of signals
absorbed by the clock at time $t+r/c$ to form a
covariant 2-sphere $S$ of radius $r$ at time $t$.
If each elementary quantum operation that occurs within
the light cones removes area $A = \pi^2 \ell_P^2/2$ from the 
two-dimensional disks $D$ in the interior of the two-sphere, and
area $ 8\pi^2 \ell_P^2/3$ from the surface of the two-sphere, the resulting 
curvature makes Einstein's equations hold.}

\label{fig:fig1}
\end{center}
\end{figure}

\end{document}